\documentclass[twocolumn,showpacs,preprintnumbers,amsmath,amssymb]{revtex4-1}

\usepackage{graphicx}
\usepackage{dcolumn}
\usepackage{bm}

\begin{document}

\preprint{APS/123-QED}

\title{Diffusion-less recrystallization at high uniaxial deformation}

\author{Leonid S. Metlov, Anatoliy V. Zavdoveev}

 \email{lsmet@fti.dn.ua, zavdoveev@fti.dn.ua}
\affiliation{Donetsk Institute of Physics and Engineering, Ukrainian
Academy of Sciences,
\\83114, R.Luxemburg str. 72, Donetsk, Ukraine
}%

\date{\today}

\begin{abstract}
It was shown by computer simulation that coarsened grains with lower content 
of defects are formed at uniaxial deformation of a four-grain infinite nano-wire. 
The structure similar to crystal filaments was formed in the case of tension. 
The case of compression demonstrated formation of two grains (from four initial one) 
disoriented at an angle of four degrees. 
\end{abstract}

\maketitle

\section{Introduction}

It is well known that the process of megaplastic (severe) plastic deformation (SPD) 
is accompanied by fragmentation (size reduction) of grains due to multiplication 
of dislocations, grain boundaries, and other defects. 
It is difficult to imagine that the same force factors allow occurrence of 
opposite diffusion-free processes directed toward grain coarsening, which can 
be considered as a specific kind of crystallization. Recently, a conception was 
developed that the processes of metal grain fragmentation and recrystallization 
periodically alternate in time \cite{g07}. 
At that, new relaxation mechanisms are activated such as low-temperature dynamical 
recrystallization \cite{g07,gm10}. It is assumed that diffusion accelerated by external 
stresses plays an important role in these processes \cite{ufb49}. However it is known that 
diffusion processes are slow enough. 
At recrystallization anneal, time of treatment is about one hour, and the time of homogenization anneal is ten hours. 
So long time periods are required for realization of structural transformations.
 Even if we suggest diffusion accelerated by an order, it is impossible to explain high rate of these processes. 
At the same time, samples of tens millimeters in size are processed with SPD methods during several
 fractions of a second.

\begin{figure}
\hspace{0.06 cm}
\includegraphics [width=3 in] {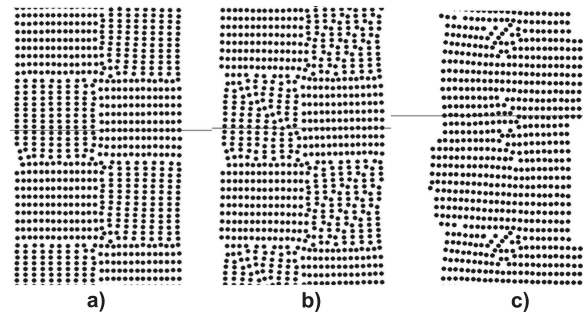}
\caption{\label{f1} Structure changes at the atomic level at uniaxial  tensile deformation: 
а) the initial state ; b, c) the deformed state; steps = 423, 3259, 20755 correspondingly.}
\end{figure}

It was shown earlier by one of the authors that methods of molecular
 dynamics predict another diffusion-free recrystallization based on the same principle as martensitic transition \cite{mbd05}. 
The experiment was carried out at a two-grain infinite nano-wire with free external boundaries at uniaxial tension,
 i.e. at conditions far from a real SPD. The experiment was organized in such a way that the boundaries of 
the external blocks had lugs, so the correlation with the real situation was complicated. 
Here the results of simulation of four-grain nano-wire are presented,
 with the wire constructed without lugs of the external free faces (Fig. \ref{f1}a). 
Both tension and compression tests were carried out. 

It was noticed by Tott that the initial texture of the sample exerts great influence on deformation 
processes \cite{telg10}. He suggested a model of grain fragmentation in metal at severe deformations, 
which was based on a combination of deformation-strengthening model and Taylor’s model with 
a new element added, i.e. possible evolution of grain population by means of fragmentation. 
Structure evolution of fcc metals in Brigeman anvils demonstrated that grains with ideal 
orientation relative to the external stress (or deformation) were much less fragmented compared 
to the grains of other orientations. 
To a certain extent, we have checked this feature in the course of computer simulation. 
In our case, we form the initial texture of double type with a half of grains oriented 
at an angle of 90 degree about the other half of grains, i.e. 50 percent of «crystallites» are ideally oriented.

\section{Calculation technique}

Numerical methods of molecular dynamics are often used for 
studying of peculiarities of the kinetics of defects at atomic layer 
\cite{fs02,m10}. Here two aspects of the problem are of special importance, 
that is generation and motion of structure defects and the related 
heat production (entropy production). 
We suggest that the particles interact with the aid of Lennard-Jones pair potential (\ref{b1}):
\begin{eqnarray}
\label{b1}
U_{ijklmn}=E_{b}((\dfrac{r_{0}}{r_{ijklmn}})^{12}-2(\dfrac{r_{0}}{r_{ijklmn}})^{6}),  \
\end{eqnarray}
where $r_{ijklmn}$ is the distance between the particles 
with the numbers of $i$, $j$, $k$ and $l$, $m$, $n$ and with Cartesian coordinates $X_{ijk}$, $Y_{ijk}$, $Z_{ijk}$ 
and $X_{lmn}$, $Y_{lmn}$, $Z_{lmn}$. 
The accepted short numeration of the particles is convenient 
for determination of the initial state of $3D$ lattice and the succeeding 
control of its evolution. Indexes $i$ and $l$, $j$ and $m$, $k$ and $n$ 
numerate atoms of the lattice along $X$, $Y$ and $Z$ direction, correspondingly. $E_{b}$, 
$r_{0}$ are the binding energy and equilibrium distance between the particles in 
a dimer, respectively. Evaluations can be applied to copper, 
so we assume $E_{b}=0.83·10^{-19} J, r_{0}=0.3615·10^{-9} m$, and $M=1.054·10^{-25} kg$. 
Time step was selected to be $Δt = 10^{-14}s$, the period of linear 
oscillations of a diatomic molecule was  $T=3·10^{-13}$s.

The calculations are conveniently done with using reduced units: 
$1m_{r} = 0.362 nm$; $1kg_{r} = 1.054•10^{-23} kg; 1s_{r} = 0.263•10^{-13}s.$ In this case, 
the parameter values used in the calculations will
correspond to copper. The binding energy measured in reduced units is (\ref{b2})
\begin{eqnarray}
\label{b2}
E_{b}=\dfrac{A}{12(r_{0})^{6}}=0.417*10^{-4}J_{r},  \
\end{eqnarray}

At $XY$ cross-section, the model consists of $4$ rectangles.
Along one side of each rectangle, $14$ atoms are located at the distance of $a_{0}$, 
and $10$ atoms are located along the other side at the distance of $a_{0}\sqrt{2}$ in such a way, 
so the rectangle sides appear approximately equal. This scheme allows us to arrange them, 
in order the maximal sides of the neighbor squares are mutually orthogonal (Fig.\ref{f1}a)
Four more squares just like these are located along $Z$ axis, with their atoms to be 
placed above the centers of the rectangles of the preceding layer at a height of $a_{0}/2$. 
Being continued periodically along $Z$ axis, such a construction results in creation of 
a stable fcc structure. To form a stable structure, two such layers are enough.

Thus, every two-layer square will contain $280$ atoms and the sum over the calculation 
cell will be $1120$ atoms. In addition to periodical conditions along $Z$ axis, periodical 
boundary conditions are imposed along the vertical $Y$ axis. Summarized over two squares, 
the period of boundary conditions along $Y$ axis will be $28a_{0}$ or $24$ atomic layers, whereas 
the period along $Z$ axis consists of two layers. The principle of the calculation is that 
the nearest neighbors over several coordination spheres are determined and the simulation 
is carried out over them. In terms of texture, the initial state of the system is 
approximated by infinite texture of aches-board type. The rate of deformation both 
at tension and compression was $10^{9}s^{-1}$.

\section{RESULTS AND DISCUSSION}
\subsection{Recrystallization at uniaxial stretching strain}

\begin{figure}
\hspace{0.06 cm}
\includegraphics [width=3 in] {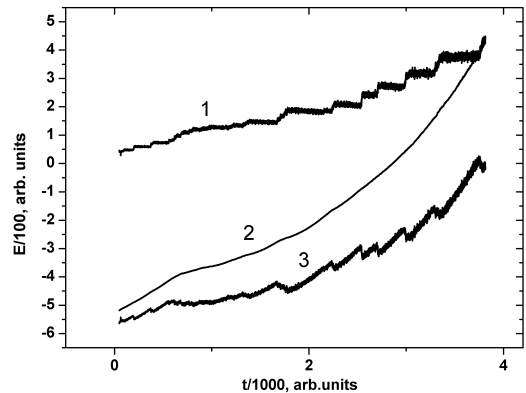}
\caption{\label{f2} Time dependence of the kinetic energy (1), the total energy (2) and the potential energy (3).}
\end{figure}

At the initial stage of every type of deformation, 
the total elastic energy of the system accumulated 
in the form of potential energy increases with deformation 
rise as the second degree according to the law of elastic 
behavior (curve 2, Fig. \ref{f2}). 
Dislocation appearance at the i-th time 
step associated with the time moment of $t = Δt•i$ results in a leap 
reduction of the potential energy. The leap lasts till the time of $t+dt$. 
The beginning of the leap is related to the appearance of a dislocation, 
and the duration is determined by dislocation motion.

Thus, the deformation of the sample of atomic size is divided into two stages: 
the elastic stage of energy accumulation and the stage of plastic motion. Further, 
these stages may alternate. The reduction of potential (elastic) energy is spent 
partially on dislocation generation (energy of inner stresses around dislocations),
 partially on the increase of the energy of heat motion at the moment of dislocation
 nucleation and during dislocation motion. This fact is rendered by dispersion of
 potential energy values at the curve.

When the plastic stage is accomplished, the tage of elastic deformation starts 
again and the potential energy increases according to a slower square relationship. 
At this stage, the leap is of smaller magnitude and lasts for a longer time period. 
The recurring leap takes place at lower level of the accumulated potential energy that 
is related to activating role of thermal fluctuations increased after the first leap.

Stage character of atomic-size sample deformation can be traced also by the measurements 
of other characteristics of the system, i.e. the total inner energy of the system 
(curve 1, \ref{f2}) and the kinetic energy (curve 3).

We should mention that the inner energy of the system is rising all the time that is 
related to the constant filling at the rate of the work done over the system. However 
the rate of the filling (incline of the plot) is not constant, being dependent on the 
stage of the process and the character of the occurring processes. After every leap the 
rate of increase of the inner energy is reduced. The plot shows also that thermal 
fluctuations of the kinetic and potential energy are strictly in antiphase That 
is expressed as absence of fluctuations of the very inner energy been evidences 
of the executed energy conservation law.

The first elastic stage of deformation begins at zero initial velocity of 
particles corresponding to the absolute zero temperature. During the first 
leap, the kinetic energy increases and then it remains stable at the succeeding 
stage of elastic deformation (curve 1 in \ref{f2}). This is an evidence of the 
absent dissipative processes in the form of dislocation motion at this stage. 
After the leap, the temperature increases slowly, being associated with both 
elastic and plastic deformation occurring simultaneously.

\begin{figure}
\hspace{0.06 cm}
\includegraphics [width=3 in] {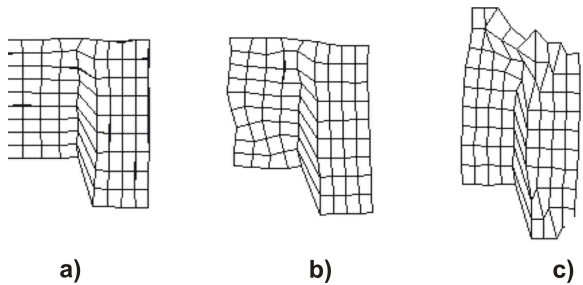}
\caption{\label{f3} Illustration of shear mechanism of structure transformation.}
\end{figure}

Under tension, a preferred direction is present in the material that is why 
the rest of crystallites tend to rearrangement along this direction. 
Crystallites are aimed to  $<100>$ orientation under tension, and to $<0-11>$ 
under compressing. Such reorientation becomes possible due to different 
mechanisms: grain boundary slip, twinning, martensitic transition. 
In our case, a transformation involving the shear mechanism (by twinning)
takes place, with the illustration presented in Fig. \ref{f3}. 
The plot presents the grid connecting atoms been neighbors at the initial 
time moment (atoms located at neighbor sites of the calculated array). 
The figure demonstrates also some lattice distortions determined by 
diffusion of separate atoms.

\begin{figure}
\hspace{0.06 cm}
\includegraphics [width=3 in] {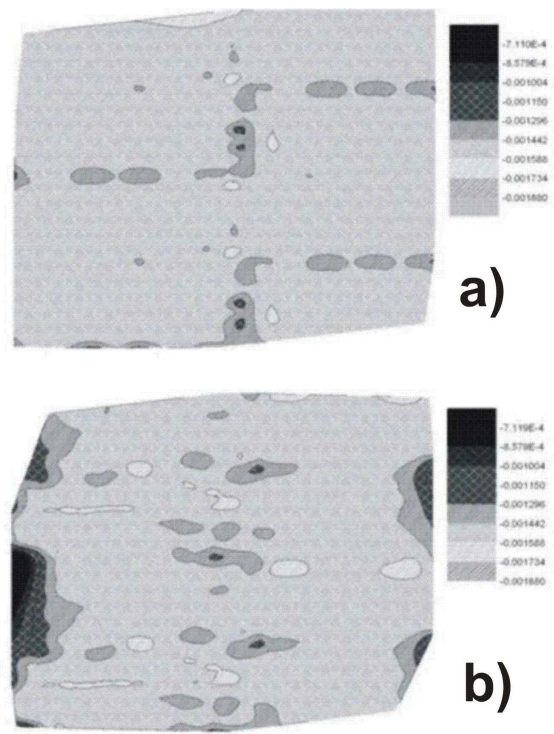}
\caption{\label{f4} Distribution of the potential energy. a) is the initial state, б) is the deformed one.}
\end{figure}

We should note that as a result of evolution, that defect ensemble 
is aimed to reduction of the excess of potential energy that is seen
 in the distribution of potential energy presented in Fig.\ref{f3}. 
It was mentioned above that deformation process consists of several stages, 
i.e. the regions of the elastic deformation are followed by the regions of the plastic one (Fig. \ref{f4}). 


\subsection{Recrystallization at compression deformation}

\begin{figure}
\hspace{0.06 cm}
\includegraphics [width=4 in]{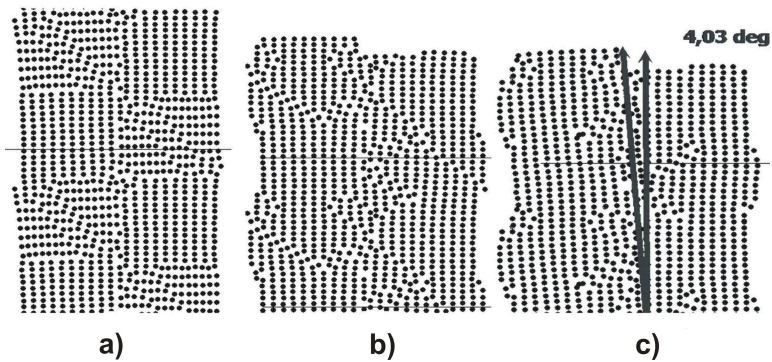}
\caption{\label{f5} Structure changes at atomic level under compression deformation: а-в – time steps $= 404, 6205, 15204, 
16703$ correspondingly. The initial stage is presented in Fig.\ref{f1}а.}
\end{figure}

An analogous situation is observed at compression deformation with some distinctions. 
Thus, uniaxial compressing forms a low-angle boundary at the final stage (Fig. \ref{f5}c),
which was not registered in the case of tension. Besides, after compression deformation, 
a texture with prevailing $<0-11>$ orientation is formed that agrees with experimental data \cite{wg62}. 
Analogous to tension, ensemble of defects tends to minimize the excess energy, 
i.e. the system trends to perfection.

Some divergences are present on the plot of potential energy, 
that is significant increase in the kinetic energy with simultaneous drop of the potential energy 
(in global sense) at the final stage of deformation. This fact is I probably related to the start 
of amorphysation (melting) of the material. Because the plot of the energy change is similar to that in 
the case of tension, we do not reproduce it. However this effect requires a more detailed study.

\subsection{Effect of deformation rate on the structural transformations of the material}

\begin{figure}
\hspace{0.06 cm}
\includegraphics [width=3 in] {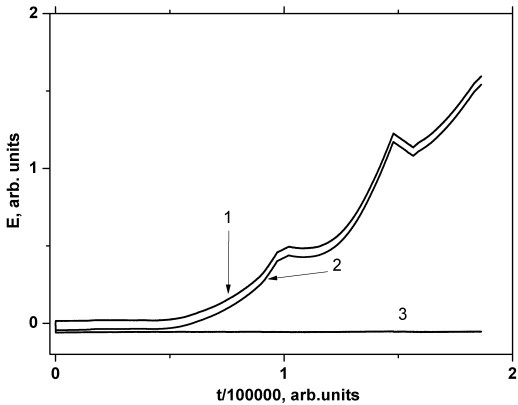}
\caption{\label{f6} STime dependence of the kinetic energy (1), the total energy (2), 
and the potential energy (3) (deformation rate is reduced by a factor of 100).}
\end{figure}

In the present work, we have also considered the cases of deceleration of the deformation. 
For instance, if the deformation rate is reduced tenfold, no distinctions in kind are observed. 
However, if the rate drops by two orders, another evolution of the system is registered.
The main distinction is that contradicting to the above theory, 
the potential energy does not change at first glance (Fig. \ref{f6}.).
But actually there are no contradictions because in the case of such slow deformation rate, the system 
is «shallowly unloaded». When speaking about «shallowly unload» we 
suggest the system accommodation occurring at lower level of the accumulated potential energy 
that is related to activating role of thermal fluctuations. Their level has increased after the 
preceding structural transformation.  Besides, global leaps of the total energy are observed, 
in contrast to high deformation rates. This effect may be determined by the work performed by the system. 
Undoubtedly, this phenomenon is worth of more detailed consideration.

As for the structure transformation, all the stages of structure reconstruction observed at the 
relatively high rate are present here. For example, the system passes the stage of twin formation,
with twins formed in grains without prevailing orientation. Further, the formation of phase synchronism occurs, 
that propagate along the whole studied subject both in vertical and horizontal direction. 
Then the system is deformed as a single whole.

\section{SHORT CONCLUSION}

It was shown that in both cases, a martensitic-type transition results in formation of coarsened grains 
with lower content of defects of smaller dimensionality as compared to grain boundaries. At tension, 
a structure of a crystal filament type was formed. Structure transformation at uniaxial compression 
was finished with formation of two grains disoriented at an angle of 4 degree (from four initial ones).

We have shown conceptual possibility of diffusion-free structure transformation using shear mechanism 
(the way of twinning) at the deformation of a nan-subject under uniaxial loading .

The effect of the deformation rate on the behavior of the system of deformed nano-grains is also 
presented. At lower deformation rates, system accommodation takes place at  smaller level of the 
accumulated potential energy.

\begin{acknowledgments}

The authors express their gratitude to Dr. of Sci. Pashinskaya E.G. and Dr. of Sci. 
Konstantinova T.E. for effective discussion and valuable advices and comments.

\end{acknowledgments}

\end{document}